\documentclass[pre, twocolumn, showkeys, showpacs, amsmath, amssymb]{revtex4}

\usepackage{graphicx}
\begin{document}

\title{A comparison of three replication strategies in complex multicellular organisms:  Asexual replication, sexual replication with identical gametes, and sexual replication with distinct sperm and egg gametes}

\author{Emmanuel Tannenbaum}
\email{emanuelt@bgu.ac.il}
\affiliation{Department of Chemistry, Ben-Gurion University of the Negev, Be'er-Sheva, Israel}

\begin{abstract}

This paper studies the mutation-selection balance in three simplified replication models.  The first model considers a population of organisms replicating via the production of asexual spores.  The second model considers a sexually replicating population that produces identical gametes.  The third model considers a sexually replicating population that produces distinct sperm and egg gametes.  All models assume diploid organisms whose genomes consist of two chromosomes, each of which is taken to be functional if equal to some master sequence, and defective otherwise.  In the asexual population, the asexual diploid spores develop directly into adult organisms.  In the sexual populations, the haploid gametes enter a haploid pool, where they may fuse with other haploids.  The resulting immature diploid organisms then proceed to develop into mature organisms.  Based on an analysis of all three models, we find that, as organism size increases, a sexually replicating population can only outcompete an asexually replicating population if the adult organisms produce distinct sperm and egg gametes.  A sexual replication strategy that is based on the production of large numbers of sperm cells to fertilize a small number of eggs is found to be necessary in order to maintain a sufficiently low cost for sex for the strategy to be selected for over a purely asexual strategy.  We discuss the usefulness of this model in understanding the evolution and maintenance of sexual replication as the preferred replication strategy in complex, multicellular organisms.

\end{abstract}



\maketitle

\section{Introduction}

\subsection{Why sex?}

The preference for sex as the exclusive mode of replication in complex multicellular life is a long-standing problem in evolutionary biology \cite{Bell:82, Bernstein:84, Williams:75, Smith:78, Hurst:96}.  The difficulty with understanding the preference for sexual replication is that, at first glance, sex appears to be a cumbersome and inefficient mode of reproduction:  While asexually replicating organisms can simply produce asexual spores that then develop into new adult organisms, sexual replication involves the mixing of genetic material from two distinct organisms.  

The need to combine genetic material from two distinct organisms incurs a fitness penalty, in the form of time and energy costs that each sexually replicating organism must pay in order to find a genetic recombination partner.  Further, in cases where a population employs a sexual replication strategy that involves distinct sexes producing distinct egg and sperm gametes, the potential rate of reproduction is half that of an asexually replicating population, where each organism produces diploid eggs.  This $ 50\% $ difference in reproduction rates is known as the {\it two-fold cost for sex} \cite{Williams:75}. 

\subsection{Theories for the existence of sex}

Despite the disadvantages for sex, its preference, which is in many cases exclusive, in the higher organismal lines indicates that it must confer an overall fitness advantage in complex organisms.  

A variety of theories have been proposed for the benefits of sex, which generally fall into one of two categories:  (1) Sex increases adaptability; and (2) Sex prevents the accumulation of deleterious mutations.

The first category of theories, originally due to Weismann, state that sex evolved because it allows small populations to adapt more quickly to changing environments \cite{Hamilton:90}.  By allowing for recombination amongst different organisms, sex potentially allows for isolated beneficial mutations to become incorporated into a single organism.  The result is that sex can potentially greatly speed up rates of adaptation.  For complex, slowly replicating organisms, or in small populations, this feature of sex can provide a significant fitness advantage.

The second category of theories, by contrast, holds that sex evolved because it prevents the accumulation of deleterious mutations.  According to this theory, sex allows organisms to discard defective genes in their own genomes and replace them with functional copies \cite{Michod:95, Stauffer:9605110, Muller:64, Felsenstein:74, Haigh:74, Bernstein:85}.  

Within each category there are a number of competing theories that are variations of a general theme.
In the context of the first category, the two most common theories are the {\it Vicar of Bray Hypothesis}, and the {\it Red Queen Hypothesis}.  

The Vicar of Bray Hypothesis simply assumes that sex allows for faster adaptation in dynamic environments.  The name is derived from an English cleric who would change his political and religious views as necessary in order to remain in office \cite{Bell:82}.  

The Red Queen Hypothesis is a somewhat more complex version of the Vicar of Bray hypothesis.  It states that sex provides a fitness advantage as a result of a constant co-evolutionary genetic ``arms race" with fast replicating and evolving parasitic organisms \cite{Stauffer:9605110, Hamilton:90}.  The name is derived from a character, the Red Queen, in the story {\it Through the Looking Glass} by Lewis Carroll.  In the story, the Red Queen states, ``It takes all the running you can do, to keep in the same place." \cite{Hamilton:90}

In the context of the second category, the two most common theories are the Genetic Repair Theory, and the Muller's Ratchet Theory.  The Genetic Repair Theory simply states that sex prevents the accumulation of deleterious mutations \cite{Michod:95, Stauffer:9605110, Bernstein:85}.  The Muller's Ratchet Theory argues that sex slows down a mutation-accumulation phenomenon in small populations known as Muller's Ratchet \cite{Muller:64, Felsenstein:74, Haigh:74} (see Figure 1).

It should be noted that the various theories for sex need not be contradictory.  That is, sex may indeed prevent the accumulation of deleterious mutations {\it and} allow for faster adaptation in dynamic environments.  The reason for this is that the prevention of accumulation of deleterious mutations is accomplished by discarding defective genes, while adaptation occurs by bringing together beneficial genes.  Since these two processes are similar, if not exactly equivalent, it is possible that both categories of theories may provide complementary, rather than rival, explanations for the existence of sex.

Furthermore, although a small population may not be a requirement for a selective advantage for sex, the selective advantage for sex may nevertheless be greater in small populations than in larger ones.

\subsection{Problems with the various theories for sex}

The four theories for the selective advantage for sex have certain difficulties that makes each of them incomplete.  The adaptability category of theories requires a dynamic environment for sex to confer a fitness advantage.  However, a number of sexually replicating organisms (sharks and crocodiles, for instance), have remained apparently unevolved for tens of millions of years in what are seemingly fairly static environments \cite{Sharks, Crocodiles, Coelacanth}.  While there may be some environmental dynamics that is difficult to detect (parasites, for instance), it is not immediately clear that a dynamic environment is a necessary condition for the emergence and maintenance of sex.

The Muller's Ratchet Theory suffers from its reliance on a small population.  This is an ill-defined term, since in one context a given population may be considered large, while in another context it may be considered small.  For example, is the current human population of approximately $ 7 \times 10^9 $ people considered small or large?  If this is a large population, then the Muller's Ratchet Theory would argue that the human population should eventually become asexual.  While this supposedly large human population has existed for much too short a time compared with macroevolutionary time scales, it is nevertheless unclear if the human population would eventually become asexual if maintained at current population size and density.

The Genetic Repair Theory is the generally accepted theory for the existence of sex, mainly because it is the one that requires the fewest assumptions.  Nevertheless, it too is problematic, for if sex prevents the accumulation of deleterious mutations, then why don't {\it all} organisms replicate exclusively sexually
\cite{Fontanari:03}?
  
The issue here is not why genetic recombination is advantageous.  It is well-known that genetic recombination between organisms occurs at all levels of organismal complexity, including bacteria and even viruses \cite{Boerlijst:96}.  Furthermore, certain single-celled organisms, such as {\it Saccharomyces cerevisiae}, or Baker's yeast, can engage in a form of sexual replication as part of a stress response \cite{Zeyl:97}.  However, the single-celled organisms that are capable of genetic recombination either replicate themselves asexually, or only replicate sexually when stressed.  This behavior is in sharp contrast to more complex, multicellular organisms, which replicate sexually either exclusively or nearly exclusively.

Clearly then, there are regimes where the advantages for sex are outweighed by its disadvantages.  A complete theory for the existence of sex must be able to identify the regimes where either sexual or asexual replication are respectively dominant, in a manner that is consistent with observation.

\subsection{Haploid fusion time and its affect on the selective advantage for sex}

\subsubsection{A density-independent model assuming first-order haploid fusion kinetics}

In \cite{Tannenbaum:06}, the author developed a mathematical model describing the evolutionary dynamics of a unicellular population that could replicate either asexually or sexually.  The sexual replication mechanism was loosely based on the sexual stress response in {\it Saccharomyces cerevisiae} (Baker's yeast) (see Figure 2).  The sexual replication model explicitly incorporated a time cost for sex, in the form of a characteristic time for haploid cells to find each other and recombine.  

What the author found was that when the haploid fusion time was small compared to the characteristic growth and doubling time of the cell, sexual replication was the advantageous strategy.  However, when the haploid fusion time was large compared to the characteristic growth and doubling time of the cell, then sexual replication was the disadvantageous strategy.  These results implied that sex should be the preferred replication strategy in slowly replicating organisms, and, all other factors being equal, in regimes of high population density.  These results therefore delineated regimes where sexual and asexual replication are respectively advantageous in a biologically consistent manner. 

\subsubsection{A density-dependent model assuming second-order haploid fusion kinetics}

The sexual replication model developed in \cite{Tannenbaum:06} assumed a density-independent haploid fusion time, an unrealistic simplifying assumption that was changed in \cite{Tannenbaum:06-2} to a second-order rate process (the assumption of a second-order rate process is a natural one to make if haploid fusion is modeled as occurring via random binary collisions between haploid cells).  While the results of this paper were qualitatively similar to the results in \cite{Tannenbaum:06}, the paper raised interesting questions regarding the emergence of sexual replication in an asexually replicating population that did not arise with the density-independent model.

\subsubsection{Multicellular organisms}

Although the work in \cite{Tannenbaum:06, Tannenbaum:06-2} suggested that sex should be favored in slowly replicating organisms, it did not work with a replication model appropriate for multicellular organisms.  With multicellular organisms replicating asexually, single-celled spores, or eggs, are continually produced by the adult organism, which may then develop into an adult organism.  

In the case of multicellular organisms replicating sexually, haploid gametes are continually produced by diploid germ cells.  These gametes may recombine with other haploid gametes, to form ``post-fusion" spores (e.g. fertilized eggs) that may then develop into an adult organism (see Figure 3).

In \cite{Tannenbaum:07}, the author developed mathematical models for asexual and sexual replication based on sporulation.  The sexual model did not assume internal fertilization or even explicit mating between organisms.  Rather, the model adopted a sexual replication model based on {\it external fertilization}, whereby haploids are released into an external aqueous environment, and fuse with other haploids in the haploid pool.  More specifically, the sexual replication model assumed a specific form of external fertilization known as {\it broadcast fertilization} or {\it spawning}.  In this fertilization mechanism, the adult organisms do not attempt to get closer to one another before releasing their gametes.  Rather, the adult organisms simply release their gametes into a general pool, so that the adult organisms do not facilitate haploid fusion in any way \cite{Randerson:01}.  This model reflects the fact that terrestrial life, both uni- and multi-cellular, developed in the oceans, and that sexual replication in the older organismal lines involves the release of haploids into the water \cite{Randerson:01}.  Therefore, the simplest models capable of identifying a selective advantage for sexual replication in complex organisms, and a disadvantage for sexual replication in simple organisms, should be able to do so with this primitive fertilization mechanism.

Assuming second-order kinetics for the haploid fusion rate, the author found that sex is indeed the preferred replication strategy when the time cost for sex is small compared to the characteristic growth time of the organism.  However, the model also suggested that as organism size increases, the consequent {\it decrease} in population density implies that the cost for sex should {\it increase} as organism size grows.  It was found that, in this case, sex can only provide a selective advantage over asexual replication if the rate of production of gametes decreases as $ 1/N $, where $ N $ is the number of cells in the adult organism.  This is clearly a problematic result, since multicellular organisms can produce gametes in great numbers (the males of some species, for instance, can produce millions of sperm every day) \cite{Moller:89}.

\subsection{Sexual replication with sperm and egg gametes}

As will be shown in this paper, it turns out that the resolution to this discrepancy rests on the following crucial assumption:  If a fertilized spore (egg) develops in an external aqueous environment, then it must contain a sufficient amount of material in order to allow the initial cell to develop into an immature multicellular organism that is capable of functioning independently of its parents.  In short, the adult organisms do not produce gametes that are of microscopic size, but rather {\it eggs} that contain a sufficient amount of extra-genetic material (the ``yolk") to allow the organism to develop sufficiently so that it has a reasonable chance of survival after hatching.  

Working from this assumption, we find that, as organism size increases, it is necessary for the adult population to produce two distinct types of gametes:  Large, relatively immobile eggs that contain the necessary non-genetic material to produce a viable immature organism, and small, highly mobile sperm that fertilize the eggs.  This type of replication strategy is in contrast to the replication strategies considered in \cite{Tannenbaum:06, Tannenbaum:06-2, Tannenbaum:07}, which assumed gametes of more or less identical size (isogamy).    

Distinct sperm and egg gametes (anisogamy) are found to be crucial for maintaining a sufficiently low cost for sex so that it is the preferred replication strategy as organism size increases.  The results of this distinct gamete model therefore provide a useful insight into the preference for a sexual replication strategy in complex multicellular life, and also suggests a resolution to the two-fold cost for sex and the necessity of males.

This paper is organized as follows:  In the following section (Section II), we review the asexual replication model considered in \cite{Tannenbaum:07} .  Since we are explicitly considering eggs in this paper, some of the terminology is slightly changed.  In Section III, we review the identical-gamete sexual replication model considered in \cite{Tannenbaum:07}.  In Section IV, we develop and analyze the distinct-gamete sexual replication model.  In Section V we compare all three replication models, and show that a distinct-gamete model is necessary for sexual replication to be advantageous over asexual replication as organism size increases.  In Section VI we further discuss the implications of our results.  Finally, in Section VII we summarize the main conclusions of this paper, and describe plans for future research.

As a concluding note for this section, it should be noted that there has been a considerable amount of research on the selective advantage for a distinct-gamete, or anisogamous, sexual replication strategy \cite{Randerson:01, Dusenbery:00, Dusenbery:02, Dusenbery:06}.  Indeed, one of the main theories for the selective advantage of an anisogamous strategy is that it reduces the time cost associated with haploid fusion.  

In one version of this theory, an anisogamous strategy reduces the time cost associated with haploid fusion because many sperm can be produced for every egg, so that the eggs are essentially bathed in a ``sperm cloud" \cite{Randerson:01}.  

In another version of this theory, an anisogamous strategy reduces the time cost associated with haploid fusion because the small sperm are highly mobile and the eggs provide a large surface area for contact, which together leads to an increase in sperm-egg contact frequency \cite{Dusenbery:06}.  

However, previous research on the evolution of anisogamy has focused on the selective advantage of a distinct-gamete sexual replication strategy over identical-gamete, or isogamous, sexual replication strategies.  This paper, by contrast, argues that, within the context of broadcast fertilization, {\it gamete differentiation is necessary in order to provide a selective advantage for the sexual strategy itself}.  Nevertheless, because the idea that distinct gametes reduce the time cost for sex has been advanced before, our arguments will be similar to those of previous authors.  In particular, this paper follows the approach of Dusenbery, and applies the kinetic theory of gases to analyze the haploid fusion process \cite{Dusenbery:06, Boltzmann:64, Kauzmann:66, Cox:85, Gerritsen:77}.

\section{The Asexual Replication Model}

\subsection{Definitions}

We begin by summarizing the asexual replication model considered in \cite{Tannenbaum:07}, and which will be used as a basis for comparison with the other replication models in this paper.  

We assume that we have a population of diploid organisms, whose genomes consist of two chromosomes.  Each chromosome may be represented as a linear symbol sequence $ \sigma = s_1 s_2 \dots s_L $, where $ L $ is the number of letters (equivalently, bases) in the sequence, and where $ s_i $ denotes the $ i^{\mbox{th}} $ letter (so $ s_i = \mbox{ A, T, G, or C} $ for DNA, A, U, G, C for RNA).  A given genome may then be represented by the set $ \{\sigma, \sigma'\} $, where $ \sigma $ and $ \sigma' $ denote the base sequences of the two chromosomes.

We also assume that there is a ``master" sequence, denoted $ \sigma_0 $, for which the chromosome is functional, and that any mutation to $ \sigma_0 $ renders the chromosome defective.  It is then assumed that the fitness of an organism (to be defined below) is determined by the number of functional chromosomes (zero, one, or two) in the organism.  

While this simplifying assumption is clearly a highly coarse approximation of organismal fitness, it is the generalization of the Single Fitness Peak Landscape commonly used in quasispecies theory \cite{Galluccio:97, Alves:98, Alves:97, Campos:98, Campos:99}.  This fitness landscape is used because it is the simplest landscape that is both analytically tractable, and that gives phenomenological results that are in qualitative agreement with a number of important biological effects (in certain cases, these results can even be quantitative) \cite{Kamp:02}.  Therefore, we regard this choice of landscape as a useful starting point for our model.  It should be mentioned, however, that recent research in quasispecies theory \cite{Eigen:89, Wilke:05, Bull:05}, and evolutionary dynamics in general, has focused on developing and analyzing more realistic fitness functions \cite{Tannenbaum:04, Shakhnovich:07}.

Within the context of our chosen fitness landscape, the population can be divided into three categories:  The first category consists of organisms with genome $ \{\sigma_0, \sigma_0\} $.  These organisms have genomes consisting of two functional chromosomes.  If we define the chromosome $ \sigma_0 $ to be {\it viable}, then we may define organisms with two viable (i.e. functional) chromosomes to be of type $ vv $.

The second category consists of organsims with genome $ \{\sigma_0, \sigma \neq \sigma_0\} $.  These organisms have genomes consisting of one functional and one non-functional chromosome.  If we define the chromosomes $ \sigma \neq \sigma_0 $ to be {\it unviable}, then we may define organisms with one viable and one unviable chromosomes to be of type $ vu $.

The third and final category consists of organisms with genome $ \{\sigma \neq \sigma_0, \sigma' \neq \sigma_0 \} $.  These organisms have genomes consisting of two non-functional chromosomes, and are defined to be of type $ uu $.

The asexual replication process, as illustrated in Figure 4, is assumed to occur as follows:  An adult organism continually generates asexual, single-celled spores at some genome-dependent rate $ \omega_{\{\sigma, \sigma' \}} $.  These spores, which we also refer to as immature organisms, develop into adult organisms with a first-order rate constant, denoted $ \kappa_{\{\sigma, \sigma'\}} $.

We define $ \omega_{vv} = \omega_{\{\sigma_0, \sigma_0\}} $, $ \omega_{vu} = \omega_{\{\sigma_0, \sigma \neq \sigma_0\}} $, and $ \omega_{uu} = \omega_{\{\sigma \neq \sigma_0, \sigma' \neq \sigma_0\}} $.  We also define $ \kappa_{vv} $, $ \kappa_{vu} $, and $ \kappa_{uu} $ in an analogous manner.

Throughout this paper, we will assume, unless otherwise stated, that, $ 0 = \omega_{uu} \leq \omega_{vu} \leq \omega_{vv} $, and $ 0 = \kappa_{uu} \leq \kappa_{vu} \leq \kappa_{vv} $.  This makes sense, since organisms with two defective chromosomes should not be expected to grow or sporulate, and organisms with one defective chromosome should not be expected to grow or sporulate more rapidly than organisms with no defective chromosomes.

Furthermore, we define $ \alpha = \kappa_{vu}/\kappa_{vv} $, and $ \beta = \omega_{vu}/\omega_{vv} $.
These parameters may be interpreted as the growth and sporulation penalties, respectively, associated with having a defective chromosome.

The actual production of spores is assumed to occur as follows:  A subset of the cells in the adult organism are responsible for producing spores.  These are the spore stem-cells, otherwise known as {\it germ cells}.  Each of these cells periodically divide to produce two daughter cells, one of which remains with the population of the spore-producing cells, while the other goes on to become a spore. 

Within this model for spore production, we assume that a given parent chromosome has a probability $ p $ of replicating correctly.  We also assume that the sequence length $ L $ is sufficiently large that the probability that an already mutated base will mutate again is negligible, so that any new mutations in the daughter strand must occur in a previously un-mutated portion of the genome.  This assumption, termed the {\it neglect of backmutations}, means that the probability of a $ v $ parent chromosome producing a $ v $ daughter chromosome is $ p $, the probability of a $ v $ parent chromosome producing a $ u $ daughter chromosome is $ 1 - p $, and the probability of a $ u $ parent chromosome producing a $ u $ daughter chromsome is $ 1 $.  
 
Finally, we assume that when a given spore-producing cell divides, the daughter cell that remains the spore-producing cell retains the parent chromosomes, while the daughter cell destined to become a spore retains the daughter chromosomes.  This chromosome segregation mechanism, termed {\it immortal strand segregation}, is known to occur in adult stem cells, and is believed to be the chromosome segregation method in budding yeast \cite{Cairns:75, Merok:02, Potten:02, Tannenbaum:05}.  Because of its apparent ubiquity in stem cell populations, we adopt this chromosome segregation method here, though it is possible to consider random recombination, as was done in \cite{Tannenbaum:06, Tannenbaum:06-2}.

Figure 5 illustrates the details of the spore-production process.  

\subsection{Population genetics equations and the mean fitness}

If we let $ n_{am, vv} $, $ n_{am, vu} $, $ n_{am, uu} $ denote the number of adult organisms with genomes $ vv $, $ vu $, and $ uu $ respectively, and if we let $ n_{ai, vv} $, $ n_{ai, vu} $, $ n_{ai, uu} $ denote the number of immature organisms (spores) with genomes $ vv $, $ vu $, $ uu $ respectively, then the time evolution of the various populations is given by the following system of ordinary differential equations:

\begin{eqnarray}
& &
\frac{d n_{am, vv}}{dt} = \kappa_{vv} n_{ai, vv}
\nonumber \\
& &
\frac{d n_{am, vu}}{dt} = \kappa_{vu} n_{ai, vu}
\nonumber \\
& &
\frac{d n_{am, uu}}{dt} = \kappa_{uu} n_{ai, uu}
\nonumber \\
& &
\frac{d n_{ai, vv}}{dt} = p^2 \omega_{vv} n_{am, vv} - \kappa_{vv} n_{ai, vv}
\nonumber \\
& &
\frac{d n_{ai, vu}}{dt} = 2 p (1 - p) \omega_{vv} n_{am, vv} + p \omega_{vu} n_{am, vu} - \kappa_{vu} n_{ai, vu}
\nonumber \\
& &
\frac{d n_{ai, uu}}{dt} = (1 - p)^2 \omega_{vv} n_{am, vv} + (1 - p) \omega_{vu} n_{am, vu}
\nonumber \\
& &
+ \omega_{uu} n_{am, uu} - \kappa_{uu} n_{ai, uu}
\end{eqnarray}

The central object of interest, that we will obtain from the equations, is the {\it mean fitness} of the population, which is time-varying and therefore denoted by $ \bar{\kappa}(t) $.  If we let $ n \equiv n_{am, vv} + n_{am, vu} + n_{am, uu} $ denote the total population of adult organisms, then we define $ \bar{\kappa}(t) = (1/n) (d n/dt) $, so that the mean fitness is simply the per-capita growth rate of the population as a whole.

The mean fitness measures the effective first-order growth rate constant of the population.  As a result, if two separate populations are placed in a given environment, the population with the higher mean fitness will drive the other to extinction.  Therefore, the mean fitness allows us to determine which replication strategy (e.g. asexual, sexual with identical gametes, sexual with distinct gametes) is dominant in a given parameter regime.

It should be noted that we are working with a {\it group selection} approach for determining which replication strategy is dominant for a given set of parameters.  While such an approach is often the easiest to analyze, one has to be careful in using it, since it is well-known from evolutionary game theory that the evolutionarily stable strategy (ESS) does not always coincide with the replication strategy that maximizes the overall fitness of the entire population \cite{Tannenbaum:06-2, Maynard-Smith:82}.  This is an issue that we will discuss toward the end of this paper.

Although in this simplified model the population can grow indefinitely, the population fractions eventually reach a steady-state, and so the mean fitness converges to a steady-state as well.  To determine the steady-state mean fitness, we define a new set of parameters that converge to steady-state as well, and re-express our dynamical equations in terms of them.

To this end, we define $ x_{aq, rs} = n_{aq, rs}/n $, where $ q = m, i $, and $ rs = vv, vu, uu $.  With these definitions, we have that,
\begin{equation}
\bar{\kappa}(t) = \kappa_{vv} x_{am, vv} + \kappa_{vu} x_{am, vu} + \kappa_{uu} x_{am, uu}
\end{equation}

Re-expressing the dynamical equations in terms of the $ x_{aq, rs} $ population ratios, we obtain,
\begin{eqnarray}
& &
\frac{d x_{am, vv}}{dt} = \kappa_{vv} x_{ai, vv} - \bar{\kappa}(t) x_{am, vv}
\nonumber \\
& &
\frac{d x_{am, vu}}{dt} = \kappa_{vu} x_{ai, vu} - \bar{\kappa}(t) x_{am, vu}
\nonumber \\
& &
\frac{d x_{am, uu}}{dt} = \kappa_{uu} x_{ai, uu} - \bar{\kappa}(t) x_{am, uu}
\nonumber \\
& &
\frac{d x_{ai, vv}}{dt} = p^2 \omega_{vv} x_{am, vv} - (\kappa_{vv} + \bar{\kappa}(t)) x_{ai, vv}
\nonumber \\
& &
\frac{d x_{ai, vu}}{dt} = 2 p (1 - p) \omega_{vv} x_{am, vv} + p \omega_{vu} x_{am, vu} 
\nonumber \\
&  &
- (\kappa_{vu} + \bar{\kappa}(t)) x_{ai, vu}
\nonumber \\
& &
\frac{d x_{ai, uu}}{dt} = (1 - p)^2 \omega_{vv} x_{am, vv} + (1 - p) \omega_{vu} x_{am, vu}
\nonumber \\
& &
+ \omega_{uu} x_{am, uu} - (\kappa_{uu} + \bar{\kappa}(t)) x_{ai, uu}
\end{eqnarray}

\subsection{Steady-state behavior}

The steady-state solution may be found by setting the left-hand-sides of these equations to zero, and solving for the various population ratios and for the steady-state mean fitness.  We obtain \cite{Tannenbaum:07},
\begin{eqnarray}
&  &
x_{ai, vv} = \frac{\bar{\kappa}(t = \infty)}{\kappa_{vv}} x_{am, vv}
\nonumber \\
&  &
x_{ai, vu} = \frac{\bar{\kappa}(t = \infty)}{\kappa_{vu}} x_{am, vu}
\nonumber \\
&  &
x_{ai, uu} = \frac{\bar{\kappa}(t = \infty)}{\kappa_{uu}} x_{am, uu}
\end{eqnarray}
which may be substituted into the bottom three equations to give,
\begin{eqnarray}
&  &
0 = [\bar{\kappa}(t = \infty)^2 + \kappa_{vv} \bar{\kappa}(t = \infty) - p^2 \omega_{vv} \kappa_{vv}] x_{am, vv}  
\nonumber \\
&  &
0 = -2 p(1 - p) \omega_{vv} \kappa_{vu} x_{am, vv} 
+ [\bar{\kappa}(t = \infty)^2 
\nonumber \\
&  &
+ \kappa_{vu} \bar{\kappa}(t = \infty) - p \omega_{vu} \kappa_{vu}] x_{am, vu}
\nonumber \\
&  &
0 = -(1 - p)^2 \omega_{vv} \kappa_{uu} x_{am, vv} - (1 - p) \omega_{vu} \kappa_{uu} x_{am, vu} 
\nonumber \\
&  &
+ [\bar{\kappa}(t = \infty)^2 + \kappa_{uu} \bar{\kappa}(t = \infty) - \omega_{uu} \kappa_{uu}] x_{am, uu}
\end{eqnarray}

We have three cases, giving rise to three distinct steady-states:

\subsubsection{Case 1:  $ x_{am, vv} > 0 $}

If $ x_{am, vv} > 0 $ at steady-state, then it is possible to see that the steady-state mean fitness must satisfy,
\begin{equation}
0 = \bar{\kappa}(t = \infty)^2 + \kappa_{vv} \bar{\kappa}(t = \infty) - p^2 \omega_{vv} \kappa_{vv}
\end{equation}
so that,
\begin{equation}
\bar{\kappa}(t = \infty) = \kappa_{a, 1} \equiv 
\frac{1}{2} \kappa_{vv} (-1 + \sqrt{1 + 4 \frac{\omega_{vv}}{\kappa_{vv}} p^2})
\end{equation}

\subsubsection{Case 2:  $ x_{am, vv} = 0 $, $ x_{am, vu} > 0 $}

If $ x_{am, vv} = 0 $ at steady-state, then the second-to-last equation reads,
\begin{equation}
0 = (\bar{\kappa}(t = \infty)^2 + \kappa_{vu} \bar{\kappa}(t = \infty) - p \omega_{vu} \kappa_{vu}) x_{am, vu} 
\end{equation}

If $ x_{am, vv} = 0 $ and $ x_{am, vu} > 0 $ at steady-state, then this implies that,
\begin{equation}
\bar{\kappa}(t = \infty) = \kappa_{a, 2} \equiv 
\frac{1}{2} \kappa_{vu} (-1 + \sqrt{1 + 4 \frac{\omega_{vu}}{\kappa_{vu}} p})
\end{equation}

\subsubsection{Case 3:  $ x_{am, vv} = x_{am, vu} = 0 $}

If $ x_{am, vv} $ and $ x_{am, vu} = 0 $ at steady-state, then the last equation reads,
\begin{equation}
0 = (\bar{\kappa}(t = \infty)^2 + \kappa_{uu} \bar{\kappa}(t = \infty) - \omega_{uu} \kappa_{uu}) x_{am, uu}
\end{equation}

Since $ x_{am, vv} + x_{am, vu} + x_{am, uu} = 1 $, then $ x_{am, vv} = x_{am, vu} = 0 \Rightarrow x_{am, uu} = 1 $, and hence the steady-state mean fitness must be given by,
\begin{equation}
\bar{\kappa}(t = \infty) = \kappa_{a, 3} \equiv
\frac{1}{2} \kappa_{uu} (-1 + \sqrt{1 + 4 \frac{\omega_{uu}}{\kappa_{uu}}})
\end{equation}

Based on a stability analysis of the steady-states, it is possible to show that $ \bar{\kappa}(t = \infty) = \max \{\kappa_{a, 1}, \kappa_{a, 2}, \kappa_{a, 3}\} $.  Using the standing assumption that $ \omega_{uu} = \kappa_{uu} = 0 $, then $ \kappa_{a, 3} = 0 $ and $ \bar{\kappa}(t = \infty) = \max \{\kappa_{a, 1}, \kappa_{a, 2}\} $.  

In Appendix A, we show that there exists a $ p_{crit} \in [\alpha \beta, \beta] $ such that the mean fitness is given by $ \kappa_{a, 1} $ for $ p > p_{crit} $, and by $ \kappa_{a, 2} $ for $ p \leq p_{crit} $ \cite{Tannenbaum:07}.  

The transition at $ p_{crit} $ is associated with a localization to delocalization transition over one of the chromosomes in each organismal genome, so that mutation-accumulation in one of the chromosomes is governed by random genetic drift.  This phenomenon is well-known in quasispecies theory, and is termed the {\it error catastrophe} \cite{Eigen:89, Wilke:05, Bull:05}.

\section{Sexual Replication with Identical Gametes}

We now turn our attention to the identical-gamete sexual replication model analyzed in \cite{Tannenbaum:07}.  Here, we assume that when an initial diploid spore is formed from the division of a spore stem-cell, it splits into two haploid gametes that then enter a haploid pool and fuse with other haploids.  The resulting diploid is a ``post-fusion" spore (i.e. a fertilized egg) that we also term a new immature organism, which then develops into a mature adult organism in a similar manner to the asexual organisms.

\subsection{The basic model}

As with the asexual replication model, we have sporulation and maturation rate constants denoted by $ \omega_{vv} $, $ \omega_{vu} $, $ \omega_{uu} $, and $ \kappa_{vv} $, $ \kappa_{vu} $, $ \kappa_{uu} $.  Because there is no reason to assume otherwise, we will assume that these values are identical to the corresponding values in the asexual population (the populations are taken to be identical in all respects except in one aspect of their replication cycles).

We model haploid fusion as a bimolecular collision reaction characterized by second-order kinetics, with a rate constant denoted $ \gamma $.  We also assume that only the $ v $ haploids can fuse with one another, since the $ u $ haploids have a single defective chromosome, and therefore they can presumably no longer take part in the replication process.  This restrictive assumption will be relaxed in future research when we consider other kinds of mating strategies (such as random mating).

Since only the $ v $ haploids can fuse with one another, this replication strategy implies that only $ vv $ genomes are present in the population.  We therefore need only consider the population of mature $ vv $ organisms, immature $ vv $ organisms, and $ v $ haploids.  If we let the numbers of these respective populations be denoted by $ n_{sm, vv} $, $ n_{si, vv} $, and $ n_v $, respectively (the ``s" stands for {\it sexual}), then, if $ V $ denotes the system volume, the evolutionary dynamics of the population is governed by the following system of ordinary differential equations:

\begin{eqnarray}
&  &
\frac{d n_{sm, vv}}{dt} = \kappa_{vv} n_{si, vv}
\nonumber \\
&  &
\frac{d n_{si, vv}}{dt} = \frac{1}{2} (\frac{\gamma}{V}) n_v^2 - \kappa_{vv} n_{si, vv}
\nonumber \\
&  &
\frac{d n_v}{dt} = 2 p \omega_{vv} n_{sm, vv} - (\frac{\gamma}{V}) n_v^2
\end{eqnarray}
where the factor of $ 1/2 $ in the second equation comes from the fact that two haploids form one immature organism, so that if $ (\gamma/V) n_v^2 $ is the rate at which haploids disappear due to fusion, then $ (1/2) (\gamma/V) n_v^2 $ is the rate at which immature organisms are produced.

Furthermore, the factor of $ 2 $ in the third equations comes from the fact that the diploid spores divide in two, so that the rate of production of haploid gametes is twice the rate of production of diploid spores.  Since a given chromosome is replicated correctly with a probability $ p $, the rate of production of viable haploids is given by $ p $ times the rate of production of haploids.

Now, we make the further assumption that the adult organisms take up a certain amount of space, so that, as the adult population grows, so does the system volume.  To this end, we define a population density of adult organisms $ \rho = n_{sm, vv}/V $, and assume that $ \rho $ is fixed.  Furthermore, we define the population ratios $ x_{si, vv} $ and $ x_v $ via $ x_{si, vv} = n_{si, vv}/n_{sm, vv} $, and 
$ x_v = n_v/n_{sm, vv} $.

Since the adult population consists only of $ vv $ organisms, it follows that $ n = n_{sm, vv} $, and so the mean fitness $ \bar{\kappa}(t) = (1/n_{sm, vv}) (d n_{sm, vv}/dt) = \kappa_{vv} x_{si, vv} $.  Re-expressing the evolutionary dynamics equations in terms of these parameters, we obtain,

\begin{eqnarray}
&   &
\frac{d x_{si, vv}}{dt} = \frac{1}{2} \gamma \rho x_v^2 - (\kappa_{vv} + \bar{\kappa}(t)) x_{si, vv}
\nonumber \\
&   &
\frac{d x_v}{dt} = 2 \omega_{vv} p - \gamma \rho x_v^2 - \bar{\kappa}(t) x_v
\end{eqnarray}

It should be noted that this version of the identical-gamete model ignores gamete death.

\subsection{Steady-state mean fitness}

Following \cite{Tannenbaum:07}, we may find the steady-state mean fitness for this model by setting the left-hand sides of the above two equations to $ 0 $.  We obtain,

\begin{equation}
x_v = \sqrt{2 \frac{\kappa_{vv}}{\gamma \rho} (1 + \frac{\bar{\kappa}(t = \infty)}{\kappa_{vv}}) 
\frac{\bar{\kappa}(t = \infty)}{\kappa_{vv}}}
\end{equation}
which may be substituted into the second equation to give, after some manipulation,
\begin{equation}
\frac{(\frac{\omega_{vv}}{\kappa_{vv}} p - (1 + \frac{\bar{\kappa}(t = \infty)}{\kappa_{vv}}) 
\frac{\bar{\kappa}(t = \infty)}{\kappa_{vv}})^2}
{(\frac{\bar{\kappa}(t = \infty)}{\kappa_{vv}})^3 (1 + \frac{\bar{\kappa}(t = \infty)}{\kappa_{vv}})} = 
\frac{1}{2} \frac{\kappa_{vv}}{\gamma \rho} 
\end{equation}

A closed form expression for the mean fitness may be found in the limit $ \kappa_{vv}/(\gamma_{vv} \rho) \rightarrow 0 $.  This corresponds to a situation where there is no cost for sex, since the characteristic growth time, $ 1/\kappa_{vv} $, is very large compared to the characteristic haploid fusion time, $ 1/(\gamma \rho) $.

In this case, the mean fitness is obtained by solving the quadratic,
\begin{equation}
(\frac{\bar{\kappa}(t = \infty)}{\kappa_{vv}})^2 + \frac{\bar{\kappa}(t = \infty)}{\kappa_{vv}} - \frac{\omega_{vv}}{\kappa_{vv}} p = 0
\end{equation}
so that,
\begin{equation}
\bar{\kappa}(t = \infty) = \frac{1}{2} \kappa_{vv} 
(-1 + \sqrt{1 + 4 p \frac{\omega_{vv}}{\kappa_{vv}}})
\end{equation}
which may be shown to be larger than $ \kappa_{a, 1} $ for $ p \in (0, 1) $, and larger than $ \kappa_{a, 2} $ as long as $ p > 0 $ and either $ \alpha < 1 $ or $ \beta < 1 $.

Therefore, when there is no cost for sex, it is the preferred replication strategy.

\subsection{Asexual versus sexual replication as a function of $ \kappa_{vv}/(\gamma \rho) $}

We may show that $ \bar{\kappa}(t = \infty)/\kappa_{vv} $ is a decreasing function of $ \kappa_{vv}/\gamma \rho $.  To see this, note that the function $ f(x, \lambda) \equiv (\lambda - (1 + x) x)^2/(x^3 (1 + x)) $ is a decreasing function of $ x $ as $ x $ increases from $ 0 $ to the value of $ x $, denoted $ x^* $, for which $ 0 = \lambda - (1 + x) x $.  As $ x $ increases from $ 0 $ to $ x^* $, the function $ f(x, \lambda) $ decreases from $ \infty $ to $ 0 $.

Therefore, as $ \kappa_{vv}/(\gamma \rho) $ increases from $ 0 $ to $ \infty $, the mean fitness $ \bar{\kappa}(t = \infty) $ decreases from its value when there is no cost for sex down to $ 0 $.  Because the mean fitness of the sexual population is larger than the mean fitness of the asexual population when there is no cost for sex, it follows that there is a unique value for $ \kappa_{vv}/(\gamma \rho) $, denoted $ (\kappa_{vv}/(\gamma \rho))_{crit} $, where the asexual and sexual mean fitnesses become equal.  Below this value of $ \kappa_{vv}/(\gamma \rho) $, the sexual mean fitness is larger, while above this value of $ \kappa_{vv}/(\gamma \rho) $, the asexual mean fitness is larger.

To determine the critical value of $ \kappa_{vv}/(\gamma \rho) $ where the asexual and sexual strategies yield identical mean fitnesses, we simply substitute the asexual mean fitness into Eq. (15), and this directly gives us our result.  

We can study the asymptotic behavior of $ (\kappa_{vv}/(\gamma \rho))_{crit} $ in both the small and large $ \omega_{vv}/\kappa_{vv} $ limits.  When $ \omega_{vv}/\kappa_{vv} $ is small, then the asexual mean fitness is given by,
\begin{equation}
\frac{\bar{\kappa}(t = \infty)}{\kappa_{vv}} = \frac{\omega_{vv}}{\kappa_{vv}} \chi p
\end{equation}
where $ \chi = p $ or $ \beta $, depending on whether the asexual mean fitness is given by $ \kappa_{a, 1} $ or $ \kappa_{a, 2} $.

In this case, keeping the lowest-order terms involving $ \omega_{vv}/\kappa_{vv} $ in the numerator and denominator of Eq. (15), we have,
\begin{equation}
(\frac{\kappa_{vv}}{\gamma \rho})_{crit} = 2 \frac{(1 - \chi)^2}{\chi^3 p} 
(\frac{\omega_{vv}}{\kappa_{vv}})^{-1} 
\end{equation}

When $ \omega_{vv}/\kappa_{vv} $ is large, then the asexual mean fitness is given by,
\begin{equation}
\frac{\bar{\kappa}(t = \infty)}{\kappa_{vv}} = \sqrt{\frac{\omega_{vv}}{\kappa_{vv}}} \chi p^{1/2}
\end{equation}
where $ \chi = p^{1/2} $ or $ (\alpha \beta)^{1/2} $, depending on whether the asexual mean fitness is given by $ \kappa_{a, 1} $ or $ \kappa_{a, 2} $.

In this case, we keep the highest-order terms involving $ \omega_{vv}/\kappa_{vv} $ in the numerator and denominator of Eq. (15), which gives,
\begin{equation}
(\frac{\kappa_{vv}}{\gamma \rho})_{crit} = 2 (1/\chi^2 - 1)^2
\end{equation}
This equation is interesting, for it says that when $ \omega_{vv}/\kappa_{vv} $ is large, then sexual replication will outcompete asexual replication as long as $ \kappa_{vv}/(\gamma \rho) < 2 (1/\chi^2 - 1)^2 $.

Figure 6 shows a plot of $ (\kappa_{vv}/(\gamma \rho))_{crit} $ as a function of $ \omega_{vv}/\kappa_{vv} $ for given values of $ \alpha $, $ \beta $, and $ p $.

\section{Sexual Replication with Distinct Sperm and Egg Gametes}

\subsection{The need for distinct gametes}

As organism size increases, the maturation time increases as well.  If we let $ N $ denote the number of cells in the adult organism, then it has been suggested that the maturation time is proportional to either $ N^{1/4} $ \cite{West:97, Demetrius:06} or $ N^{1/3} $ \cite{Weitz:01, Weitz:06}.  To account for both possibilities, we take the maturation time to scale as $ N^{\alpha} $, where $ \alpha = 1/4, 1/3 $.  This then implies that $ \kappa_{vv} \propto N^{-\alpha} $.

If each cell takes up a volume $ \nu $, then the total volume of an adult organism is $ N \nu $.  Therefore, if there are $ n $ adults in the population, then the total volume is on the order of $ n N \nu $, so that the population density is on the order of $ 1/(N \nu) \propto 1/N $.  So, $ \rho \propto 1/N $, which gives that $ \kappa_{vv}/(\gamma \rho) \propto N^{1 - \alpha} $.  This goes to $ \infty $ as $ N \rightarrow \infty $.  

Now, for the sexual strategy to outperform the asexual strategy as organism size increases, it is necessary that, as $ N \rightarrow \infty $, $ (\kappa_{vv}/(\gamma \rho))_{crit} > \kappa_{vv}/(\gamma \rho) $.  Since $ \kappa_{vv}/(\gamma \rho) \rightarrow \infty $ as $ N \rightarrow \infty $, it follows that
$ (\kappa_{vv}/(\gamma \rho))_{crit} \rightarrow \infty $ as $ N \rightarrow \infty $.  Since this can only occur in the small $ \omega_{vv}/\kappa_{vv} $ region, it follows that we want,
\begin{equation}
\frac{\omega_{vv}}{\kappa_{vv}} < 2 \frac{(1 - \chi)^2}{\chi^3 p} \frac{\gamma \rho}{\kappa_{vv}}
\Rightarrow
\omega_{vv} < 2 \frac{(1 - \chi)^2}{\chi^3 p} \gamma \rho
\end{equation}
so that, since $ \gamma \rho \propto 1/N $, it follows that $ \omega_{vv} $ should decrease at least as quickly as $ 1/N $.  

Therefore, for sexual replication to outcompete asexual replication as organism size increases, the organismal sporulation rate should decrease as at least the reciprocal of organism size.  This is clearly too small compared with actual organismal sporulation rates, so that the identical-gamete sexual replication model presented in this section is insufficient to account for the observation that complex multicellular life prefers a sexual replication strategy (and relies on it exclusively in certain cases).

Of course, this highly simplified analysis ignores the connection between organism size and organism complexity, and hence the dependence of genome size, and of $ p $ and $ \chi $, on $ N $.  
Nevertheless, it is interesting that one does not readily obtain from a simplified model that sexual replication should be the preferred replication strategy as organism size and complexity increases.
As will be subsequently shown, the assumption of distinct gametes provides a simple solution to this problem.

\subsection{The basic model}

We consider a sexually replicating population of organisms that produce two distinct types of gametes.  We assume that one of the gametes is much smaller than the other.  The larger gamete, which we term the egg, contains, in addition to its haploid complement of chromosomes, all of the necessary material to sustain the growing embryo during its development from a single cell into an immature organism.  The smaller gamete, which we term the sperm, contains little more than the haploid complement of chromosomes.  However, the smaller size of the sperm means that they can be produced at a far greater rate than the eggs.  

We will also assume that a fraction $ \lambda $ of the population produces sperm, and a fraction $ 1 - \lambda $ produces eggs, at any given time.  This assumption does not necessarily assume a male-female split, since hermaphrodites can divide their investment in the production of sperm and eggs such that the total output corresponds to an effective fraction $ \lambda $ of the population producing sperm, and the remaining fraction $ 1 - \lambda $ of the population producing eggs.

To adapt our notation to this distinct gamete model, we let $ n_{s, v} $, $ n_{e, v} $ denote the number of viable sperm and egg haploids, respectively.  We also let $ \omega_{s, vv} $, $ \omega_{e, vv} $ denote the production rate of diploid spores that then proceed to develop into sperm and egg gametes, respectively.  All other terms are unchanged.

We also assume, for simplicity, that the sperm cells do not have any flagella or any other mechanism to allow them to actively transport themselves through the water.  Rather, we assume that the sperm are dispersed throughout the aqueous medium by convection and Brownian motion.  We will discuss this assumption, as well as a number of others, toward the end of the paper.

Finally, in contrast to the identical-gamete sexual replication model presented in the previous section, here we will generalize our model somewhat and allow for haploid death.  To this end, we let $ \kappa_{d, s} $, $ \kappa_{d, e} $ denote the first-order decay constants of the sperm and egg, respectively.

With these definitions in hand, the ordinary differential equations governing the evolutionary dynamics of the population is given by,
\begin{eqnarray}
&  &
\frac{d n_{sm, vv}}{dt} = \kappa_{vv} n_{si, vv}
\nonumber \\
&  &
\frac{d n_{si, vv}}{dt} = (\frac{\gamma}{V}) n_{s, v} n_{e, v} - \kappa_{vv} n_{si, vv}
\nonumber \\
&  &
\frac{d n_{s, v}}{dt} = 2 \omega_{s, vv} \lambda p n_{sm, vv} - (\frac{\gamma}{V}) n_{s, v} n_{e, v} - \kappa_{d, s} n_{s, v} 
\nonumber \\
&  &
\frac{d n_{e, v}}{dt} = 2 \omega_{e, vv} (1 - \lambda) p n_{sm, vv} - (\frac{\gamma}{V}) n_{s, v} n_{e, v} - \kappa_{d, e} n_{e, v}
\nonumber \\
\end{eqnarray}

If we make the same assumptions about organism density and work with the same mean fitness and population ratios as the previous sexual model, then we have,

\begin{eqnarray}
&   &
\frac{d x_{si, vv}}{dt} = \gamma \rho x_{s, v} x_{e, v} - (\kappa_{vv} + \bar{\kappa}(t)) x_{si, vv}
\nonumber \\
&   &
\frac{d x_{s, v}}{dt} = 2 \lambda \omega_{s, vv} p - \gamma \rho x_{s, v} x_{e, v} - (\kappa_{d, s} + \bar{\kappa}(t)) 
x_{s, v}
\nonumber \\
&   &
\frac{d x_{e, v}}{dt} = 2 (1 - \lambda) \omega_{e, vv} p - \gamma \rho x_{s, v} x_{e, v} - (\kappa_{d, e} + \bar{\kappa}(t)) x_{e, v}
\nonumber \\
\end{eqnarray}

In Appendix B, we will discuss how this model is affected if we assume that the egg, sperm, and immature organisms themselves take up a certain volume.

\subsection{Steady-state mean fitness}

Using the fact that $ x_{si, vv} = \bar{\kappa}(t)/\kappa_{vv} $, we have, at steady-state, that,
\begin{eqnarray}
&  &
\frac{\gamma \rho}{\kappa_{vv}} x_{s, v} x_{e, v} = (1 + \frac{\bar{\kappa}(t = \infty)}{\kappa_{vv}})
\frac{\bar{\kappa}(t = \infty)}{\kappa_{vv}}
\nonumber \\
&  &
x_{s, v} = \frac{2 \lambda \frac{\omega_{s, vv}}{\kappa_{vv}} p - (1 + \frac{\bar{\kappa}(t = \infty)}{\kappa_{vv}})
\frac{\bar{\kappa}(t = \infty)}{\kappa_{vv}}}{\frac{\bar{\kappa}(t = \infty)}{\kappa_{vv}} + 
\frac{\kappa_{d, s}}{\kappa_{vv}}}
\nonumber \\
&  &
x_{e, v} = \frac{2 (1 - \lambda) \frac{\omega_{e, vv}}{\kappa_{vv}} p - (1 + \frac{\bar{\kappa}(t = \infty)}{\kappa_{vv}})
\frac{\bar{\kappa}(t = \infty)}{\kappa_{vv}}}{\frac{\bar{\kappa}(t = \infty)}{\kappa_{vv}} +
\frac{\kappa_{d, e}}{\kappa_{vv}}}
\end{eqnarray}

The bottom two equations may be plugged into the first and re-arranged to give,
\begin{widetext}
\begin{equation}
\frac{(2 \lambda \frac{\omega_{s, vv}}{\kappa_{vv}} p - (1 + \frac{\bar{\kappa}(t = \infty)}{\kappa_{vv}})
\frac{\bar{\kappa}(t = \infty)}{\kappa_{vv}})
(2 (1 - \lambda) \frac{\omega_{e, vv}}{\kappa_{vv}} p - (1 + \frac{\bar{\kappa}(t = \infty)}{\kappa_{vv}})
\frac{\bar{\kappa}(t = \infty)}{\kappa_{vv}})}
{(\frac{\bar{\kappa}(t = \infty)}{\kappa_{vv}} + \frac{\kappa_{d, s}}{\kappa_{vv}})
 (\frac{\bar{\kappa}(t = \infty)}{\kappa_{vv}} + \frac{\kappa_{d, e}}{\kappa_{vv}})
 \frac{\bar{\kappa}(t = \infty)}{\kappa_{vv}} (1 + \frac{\bar{\kappa}(t = \infty)}{\kappa_{vv}})}
= 
\frac{\kappa_{vv}}{\gamma \rho} 
\end{equation}
\end{widetext}

This equation may be solved to give the steady-state mean fitness of the population.  In particular, when there is no cost for sex, i.e. when $ \kappa_{vv}/(\gamma \rho) \rightarrow 0 $, then we obtain that the steady-state mean fitness must satisfy one of the following equations:
\begin{eqnarray}
&  &
(\frac{\bar{\kappa}(t = \infty)}{\kappa_{vv}})^2 + (\frac{\bar{\kappa}(t = \infty)}{\kappa_{vv}}) - 
2 \lambda \frac{\omega_{s, vv}}{\kappa_{vv}} p = 0
\nonumber \\
&  &
(\frac{\bar{\kappa}(t = \infty)}{\kappa_{vv}})^2 + (\frac{\bar{\kappa}(t = \infty)}{\kappa_{vv}}) - 
2 (1 - \lambda) \frac{\omega_{e, vv}}{\kappa_{vv}} p = 0
\nonumber \\
\end{eqnarray}

Because in general $ \omega_{e, vv} << \omega_{s, vv} $, we expect that $ \lambda \omega_{e, vv} << (1 - \lambda) \omega_{s, vv} $.  This implies that the second equation yields the smaller steady-state mean fitness, and so this is the steady-state mean fitness of the population (essentially, because the eggs are released at a lower rate than the sperm, the eggs are the rate-limiting factor controlling fitness).

We then obtain,
\begin{equation}
\bar{\kappa}(t = \infty)= \frac{1}{2} \kappa_{vv}
(-1 + \sqrt{1 + 8 (1 - \lambda) \frac{\omega_{e, vv}}{\kappa_{vv}} p})
\end{equation}

\section{Comparison of the Various Replication Mechanisms}

\subsection{Parameter behavior in the limit $ N \rightarrow \infty $}

To show how a distinct-gamete sexual replication strategy allows for sexual replication to be the preferred replication strategy in large, complex organisms, we need to first explore how the various parameters in our models scale with organism size.  The parameters we will consider are (1) Asexual spore production; (2) Identical-gamete spore production; (3) Egg production; (4) Sperm production; 
(5) The coupling parameter $ \gamma $.

We consider each of these parameters in turn:

\subsubsection{Asexual spore production}

Complex, multicellular organisms contain genomes that encode for a multicellular survival strategy.  This survival strategy is one that should be more or less optimized for exploiting a certain environmental niche, otherwise, natural selection would eliminate it.  As a result, until the organism reaches its mature, adult size, it is not at maximal fitness.  Furthermore, because the multicellular organism is a highly interconnected network of differentiated cells, the organism must contain a minimum number of cells before it can even implement its multicellular survival strategy.  Until the organism reaches this stage of development, it is not able to exploit the niche for which it was designed, nor can it optimally function in other niches, since its genome was not designed for them. 

To illustrate, a single human fertilized egg, although it is an undifferentiated cell, is far less suited to surviving in the wild than say, a paramecium, which was specifically designed for this purpose.  The reason for this is that the fertilized egg contains numerous genes involved in implementing a multicellular survival strategy, genes that are completely unnecessary for a free-living cell.  These genes code for various biochemical pathways that cost time and energy, and do not contribute directly to the production of new cells.  As this embryo develops in the womb, it becomes even more vulnerable, for it is now in a transition from a single cell to a multi-celled organism.  During this intermediate stage, the developing human is neither able to function as a free-living cell, nor as a multi-celled human organism.

Because an organism is vulnerable until it has reached a minimal stage of development, complex organisms must provide a protected environment for a single-celled spore that allows it to develop into an immature organism that can then exploit the niche for which it was designed.  This protected environment must include sufficient quantities of pre-processed nutrients to allow the organism to develop to a certain size and complexity (so that the organism does not have to procure resources from niches for which it is poorly designed), as well as provide a defense against harsh environmental conditions, predators, and pathogens.  In Appendix C we provide a simplified analysis of these assumptions.

For many organisms, the way to provide such a protective, nutrient-filled environment is to produce relatively large eggs that contain an outer protective shell, as well as a ``yolk" that provides sufficient nutrients for the initial, single-celled spore to develop into an immature, yet fully differentiated organism that can exploit the environmental niche for which its genome was designed.  

In models for the growth of an organism to full size, which have shown a power-law dependence of growth time with the number of cells in the organism, the underlying assumption is that the metabolic rate of the organism is diffusion-limited.  The $ 1/3 $ power-law comes from assuming a spherical model, while the $ 1/4 $ power-law comes from assuming a fractal geometry to the organism, consisting of many branching blood vessels spreading throughout the body \cite{West:97, Demetrius:06}.

In any event, assuming that an asexually replicating adult organism is fully devoted to producing eggs, then if $ f $ denotes the ratio of the number of cells in the immature organism to the number of cells in the adult, we obtain that egg production may be modeled as a process whereby an $ N $-celled adult grows to $ (1 + f) N $ cells, and then the material from the $ f N $ cells are discharged to produce an egg (in reality, these $ f N $ cells consist of the ``yolk" material necessary to produce the $ f N $ - celled organism later on).

Assuming that the rate of growth of the adult is proportional to the metabolic rate, with a constant of proportionality $ \eta $, that the metabolic rate scales as $ N^{1 - \alpha} $, and that $ f $ is small, we obtain,
\begin{equation}
\tau_{egg} = \frac{f}{\eta} N^{\alpha}
\end{equation}
where $ \tau_{egg} $ denotes the characteristic time it takes to produce a single egg.

If we let $ \omega_{ae, vv} $ denote the production rate of diploid eggs in the asexual population, then we have that $ \omega_{ae, vv} \propto 1/\tau_{egg} $, so that,
\begin{equation}
\omega_{ae, vv} \propto N^{-\alpha}
\end{equation}

It should be noted that there is a class of organisms, which are termed {\it viviparous}, that do not produce relatively large eggs that are released into the external environment, but rather give birth to live offspring.  In these organisms, the eggs are much smaller, containing little to no yolk.  Instead, the organisms develop inside the female for a certain period, and then emerge as immature organisms that continue to develop on their own.

Although this paper considers broadcast fertilization, we argue that the existence of viviparous organisms does not contradict our basic argument for the existence of eggs.  In viviparous organisms, the female can afford to produce eggs with little to no yolk, because she uses her body as the protective environment in which her offspring can develop.  Furthermore, she directly provides her offspring with the nutrients necessary to grow and develop.  Therefore, the total required investment of resources in viviparous organisms is similar to that of oviparous (egg-laying) organisms (indeed, it is likely that the resource cost in viviparous organisms is higher, since the female must carry the offspring inside of her, which limits her mobility).  As a result, both viviparous and oviparous organisms have similar constraints on the rates at which females can produce offspring, leading to similar scaling of offspring production rates with size.

We should note that, if we consider parental care, then this may be regarded as an additional input of resource that further reduces the effective reproduction rate of females.

\subsubsection{Identical-gamete spore production}

With the identical-gamete sexual replication model, the organisms produce haploid eggs which must then recombine with one another to form a diploid, ``fertilized" egg that then develops into an immature organism.

We note first that if a diploid egg contains enough material to produce $ f N $ cells in the immature adult, then each haploid egg contains enough material to produce $ f N/2 $ cells.

There are two ways to produce the haploid eggs:  (1) The full-sized diploid egg divides in two, in which case the value of $ \omega_{vv} $ for this model, which we denote $ \omega_{ie, vv} $, is simply $ \omega_{ae, vv} $ (where ``i" stands for ``identical"); (2)  An initial diploid cell divides into two haploids, and each haploid is sequentially filled with enough material to form the two haploid eggs.  When $ f $ is small, then the previous analysis also yields that $ \omega_{ie, vv} = \omega_{ae, vv} $.  

Therefore, in any event we may assume that $ \omega_{ie, vv} = \omega_{ae, vv} $.

\subsubsection{Egg production}

With the distinct-gamete sexual replication model, the egg-producing organisms produce haploid eggs that are the same size as the diploid eggs in the asexual organisms, since the sperm contain essentially 
only genetic material, and no additional material to supply the growth and development of the embryo.

Following a similar analysis to the one carried out above, we may then show that, if $ f $ is taken to be small, then $ \omega_{e, vv} = \omega_{ae, vv}/2 $.  The factor of two comes from the fact that the haploid egg in this case must contain twice the material of the haploid eggs in the identical-gamete case, which reduces the egg production rate by a factor of two.

\subsubsection{Sperm production rate}

Since sperm do not contain any ``yolk" material, then, assuming that an adult sperm-producing organism is fully devoted to producing sperm, we may take $ \omega_{s,vv} $ to be proportional to the metabolic rate of the organism.  Therefore, we may assume that $ \omega_{s, vv} $ scales as $ N^{1 - \alpha} $.  For generality, however, we take $ \omega_{s, vv} \propto N^{\beta} $, where $ \beta \in (0, 1) $.  

\subsubsection{The coupling constant $ \gamma $}

The second-order rate constant that determines the haploid fusion rate depends on the size of the haploids.  On the one hand, larger haploids have a larger surface area available for coupling.  On the other hand, larger haploids will have a slower rate of random motion through the aqueous environment, which reduces the fusion rate.

Following the approach of Dusenbery \cite{Dusenbery:06}, we model haploid fusion as a binary collision process between randomly moving particles in a medium.  Since in this paper we do not assume that the sperm have any kind of active transport mechanism (e.g. flagella), the random motion is due to Brownian motion. 

Based on the theory of chemical reaction kinetics, it is possible to show that,
\begin{equation}
\gamma = \pi (r_1 + r_2)^2 \frac{3 v_1^2 + v_2^2}{3 v_1}
\end{equation}
where $ r_1 $, $ v_1 $ denote the radius and average speed of the small particles, and $ r_2 $, $ v_2 $ denote the radius and average speed of the large particles.

In the identical-gamete model, $ r_1 = r_2 = r_e $, and $ v_1 = v_2 = v_e $, where $ r_e $ and $ v_e $ are the radius and average velocity of the eggs.  This gives,
\begin{equation}
\gamma_{ig} = \frac{16 \pi}{3} r_e^2 v_e
\end{equation}
where ``ig" is meant to signify {\it identical gamete}.

Now, the radius of the egg, $ r_e $, is related to its volume via $ V_e = (4/3) \pi r_e^3 $, and the egg volume is proportional to $ f N/2 $.  Therefore, $ r_e \propto N^{1/3} \Rightarrow r_e^2 \propto N^{2/3} $.

Regarding $ v_e $, we note that diffusive motion of the eggs through the aqueous medium occurs because of random collisions of water molecules against the egg surface.  This leads to a random stochastic force $ \vec{F}(t) $, with an average magnitude $ F $.  If we assume that velocities are sufficiently small so that we are in the laminar-flow regime, then for a spherical egg the drag depends on velocity via $ D = 6 \pi \mu r_e v_e $, where $ \mu $ is the fluid viscosity \cite{Dusenbery:06}.  Equating $ F $ and $ D $ gives a velocity,
\begin{equation}
v_e = \frac{F}{6 \pi \mu r_e}
\end{equation}
and so $ v_e $ is proportional to $ 1/r_e $, and hence $ v_e \propto N^{-1/3} $.  Putting everything together, we obtain that,
\begin{equation}
\gamma_{ig} \propto N^{1/3}
\end{equation}

In the distinct-gamete model, we have $ r_1 = r_s $, $ r_2 = r_e $, where $ r_s $ is the radius of the sperm, and $ r_e $ is the radius of the eggs.  We also have $ v_1 = v_s $, and $ v_2 = v_e $.  Since the eggs are much larger than the sperm, we have $ r_s \approx 0 $, $ v_e \approx 0 $, so that,
\begin{equation}
\gamma_{dg} = \pi r_e^2 v_s
\end{equation}
where ``dg" is meant to signify {\it distinct gamete}.

Since the sperm remain more or less of constant size as organism size increases, the only quantity that scales with $ N $ is $ r_e $, giving,
\begin{equation}
\gamma_{dg} \propto N^{2/3}
\end{equation}

\subsection{How parameter behavior affects the mean fitness as $ N \rightarrow \infty $}

\subsubsection{Identical gametes}

In the identical gamete model, we have that $ \kappa_{vv}/(\gamma \rho) $ scales as
$ N^{2/3 - \alpha} \rightarrow \infty $ as $ N \rightarrow \infty $, since $ \alpha = 1/4, 1/3 $.  This implies that $ \omega_{ie, vv} $ must decrease at least as quickly as $ \gamma \rho \propto N^{-2/3} $.
Since $ \omega_{ie, vv} = \omega_{ae, vv} \propto N^{-\alpha} $, where $ \alpha = 1/4, 1/3 $, it appears that the actual decrease in egg production rates as a function of organism size is not sufficiently rapid for the sexual replication strategy to outcompete the asexual one.

\subsubsection{Distinct gametes}

In the distinct gamete model, we have that $ \kappa_{vv}/(\gamma \rho) $ scales as
$ N^{1/3 - \alpha} $.  Therefore, $ \kappa_{vv}/(\gamma \rho) $ either remains essentially constant as $ N \rightarrow \infty $, or scales as $ N^{1/12} $.  However, because we have distinct sporulation rates $ \omega_{s, vv} $ and $ \omega_{e, vv} $, we cannot directly apply the previous analysis to this model.  

To determine how the mean fitness scales as a function of organism size, we first re-write Eq. (26) as,
\begin{eqnarray}
&  &
2 (1 - \lambda) \frac{\omega_{e, vv}}{\kappa_{vv}} p - 
(1 + \frac{\bar{\kappa}(t = \infty)}{\kappa_{vv}}) \frac{\bar{\kappa}(t = \infty)}{\kappa_{vv}} 
 = 
\nonumber \\
&  &
\frac{1}{\gamma \rho \omega_{s, vv}} \frac{1}{2 \lambda p} 
\frac{\bar{\kappa}(t = \infty)}{\kappa_{vv}} (1 + \frac{\bar{\kappa}(t = \infty)}{\kappa_{vv}})
\nonumber \\
&  &
\times
(\bar{\kappa}(t = \infty) + \kappa_{d, s})(\bar{\kappa}(t = \infty) + \kappa_{d, e})
\nonumber \\
&  &
\times
\frac{2 \lambda \frac{\omega_{s, vv}}{\kappa_{vv}} p}{2 \lambda \frac{\omega_{s, vv}}{\kappa_{vv}} p - (1 + \frac{\bar{\kappa}(t = \infty)}{\kappa_{vv}}) \frac{\bar{\kappa}(t = \infty)}{\kappa_{vv}}}
\end{eqnarray}

The normalized mean fitness $ \bar{\kappa}(t = \infty)/\kappa_{vv} $ is a decreasing function of 
$ \kappa_{vv}/(\gamma \rho) $, and so has a maximal value given by Eq. (28).  We therefore have that,
\begin{eqnarray}
&  &
\frac{\bar{\kappa}(t = \infty)}{\kappa_{vv}} (1 + \frac{\bar{\kappa}(t = \infty)}{\kappa_{vv}}) \leq
2 (1 - \lambda) \frac{\omega_{e, vv}}{\kappa_{vv}} p
\nonumber \\
&  &
\bar{\kappa}(t = \infty) \leq 2 (1 - \lambda) \omega_{e, vv} p
\nonumber \\
&  &
\frac{2 \lambda \frac{\omega_{s, vv}}{\kappa_{vv}} p}
{2 \lambda \frac{\omega_{s, vv}}{\kappa_{vv}} p - 
(1 + \frac{\bar{\kappa}(t = \infty)}{\kappa_{vv}})
\frac{\bar{\kappa}(t = \infty)}{\kappa_{vv}}}
\leq 
\frac{\lambda \omega_{s, vv}}{\lambda \omega_{s, vv} - (1 - \lambda) \omega_{e, vv}}
\nonumber \\
\end{eqnarray}
These inequalities may be plugged into Eq. (37) to give,
\begin{eqnarray}
&  &
0 
\nonumber \\
&  &
\leq 
2 (1 - \lambda) \frac{\omega_{e, vv}}{\kappa_{vv}} p 
- (1 + \frac{\bar{\kappa}(t = \infty)}{\kappa_{vv}}) \frac{\bar{\kappa}(t = \infty)}{\kappa_{vv}}
\nonumber \\
&  &
\leq 
\frac{\omega_{e, vv}}{\gamma \rho \kappa_{vv} \omega_{s, vv}}
(\frac{1}{\lambda} - 1) 
\nonumber \\
&  &
\times
(2 (1 - \lambda) \omega_{e, vv} p + \kappa_{d, s})
(2 (1 - \lambda) \omega_{e, vv} p + \kappa_{d, e})
\nonumber \\
&  &
\times
\frac{1}{1 - (\frac{1}{\lambda} - 1) \frac{\omega_{e, vv}}{\omega_{s, vv}}}
\end{eqnarray}

Now, as $ N \rightarrow \infty $, $ \omega_{e, vv} $ and $ \kappa_{vv} $ scale as $ N^{-\alpha} $, $ \omega_{s, vv} $ scales as $ N^{\beta} $, $ \gamma $ scales as $ N^{2/3} $, and $ \rho $ scales as $ N^{-1} $.  This implies that $ 1/(1 - (1/\lambda - 1) (\omega_{e, vv}/\omega_{s, vv})) \rightarrow 1 $ as $ N \rightarrow \infty $.  Furthermore, if we assume that $ \kappa_{d, s}, \kappa_{e, s} > 0 $, then
$ (2 (1 - \lambda) \omega_{e, vv} p + \kappa_{d, s}) (2 (1 - \lambda) \omega_{e, vv} p + \kappa_{e, s}) \rightarrow \kappa_{d, s} \kappa_{e, s} $ as $ N \rightarrow \infty $.

Therefore, as $ N \rightarrow \infty $, the right-most term in Eq. (39) scales as $ N^{1/3 - \beta} $, which goes to $ 0 $ as long as $ \beta > 1/3 $.  In this case, the cost for sex disappears as organism size increases, so that the steady-state mean fitness is given by the mean fitness when there is no cost for sex.
  
The assumption of $ \beta > 1/3 $ certainly holds if $ \beta = 1 - \alpha $, for then we obtain that $ \beta = 2/3, 3/4 > 1/3 $.  This lower bound seems to be far below the actual rate at which the fraction of sperm-producing cells in an organism scales with organism size.  To illustrate with a simple order of magnitude calculation, the number of sperm in the average ejaculate of an adult human male is approximately $ 2.5 \times 10^{8} $ \cite{Ejaculate}, while the number of sperm in the average ejaculate of an elephant is approximately $ 2 \times 10^{11} $ \cite{Howard:84}.  If an average adult human male has a mass of $ 100 \mbox{ kg} $, while an average adult elephant male has a mass of $ 5,000 \mbox{ kg} $ \cite{Elephant}, then the average ejaculate of an adult elephant should contain $ 1.25 \times 10^{10} $ sperm.  

Therefore, the adult elephant seems to produce an order of magnitude more sperm than what would be expected assuming only a linear scaling law, suggesting that $ \beta > 1 $, at least for this example.
While we have not taken into account issues such as frequency of ejaculation, this result nevertheless suggests that the requirement that $ \beta > 1/3 $ is easily satisfied for complex organisms. 

\subsection{Comparison of the three replication mechanisms}

As $ N \rightarrow \infty $, the identical-gamete sexual replication strategy becomes too costly for it to remain competitive with either the asexual or the distinct-gamete sexual replication strategies.  However,  under the very loose requirement that $ \beta > 1/3 $, the cost for sex in the distinct-gamete strategy decreases to zero as $ N \rightarrow \infty $, so that the steady-state mean fitness of the distinct-gamete strategy for large $ N $ is given by,
\begin{equation}
\frac{\bar{\kappa}(t = \infty)}{\kappa_{vv}} = 
\frac{1}{2} (-1 + \sqrt{1 + 8 (1 - \lambda) \frac{\omega_{e, vv}}{\kappa_{vv}} p})
\end{equation}

The elimination of the cost for sex is due to the sperm gametes.  Because they are small and can be produced in large quantities compared to the eggs, for large organisms the eggs essentially become enveloped in a ``sperm cloud" that ensures rapid fertilization \cite{Hurst:96, Randerson:01}.

Using the fact that $ \omega_{e, vv} = \omega_{ae, vv}/2 $, we have that the asexual mean fitness is given by,
\begin{eqnarray}
\frac{\bar{\kappa}_{a}(t = \infty)}{\kappa_{vv}} = 
\max \{
&  &
\frac{1}{2} (-1 + \sqrt{1 + 4 \frac{\omega_{ae, vv}}{\kappa_{vv}} p^2}),
\nonumber \\
&  &
\frac{1}{2} (-\alpha + \sqrt{\alpha^2 + 4 \alpha \beta \frac{\omega_{ae, vv}}{\kappa_{vv}} p}) \}
\nonumber \\
\end{eqnarray}
while the sexual mean fitness is given by,
\begin{equation}
\frac{\bar{\kappa}_{s}(t = \infty)}{\kappa_{vv}} = 
\frac{1}{2} (-1 + \sqrt{1 + 4 (1 - \lambda) \frac{\omega_{ae, vv}}{\kappa_{vv}} p})
\end{equation}

Note then that if $ p < 1 - \lambda $, the sexual mean fitness outcompetes the asexual mean fitness when the asexual mean fitness is given by the first fitness function.  Also, if $ \beta < 1 - \lambda $, then the sexual mean fitness outcompetes the asexual mean fitness when the asexual mean fitness is given by the second fitness function.

Therefore, given values for $ p $ and $ \beta $ that are less than $ 1 $, we can choose a $ \lambda \in (0, 1 - \max \{p, \beta\}) $, and obtain that sexual replication will outcompete asexual replication for sufficiently large organisms.  Of course, the smaller the required value of $ \lambda $, the fewer the number of organisms that are involved in producing sperm, and so the larger $ N $ has to be before there is a sufficient amount of sperm to reduce the time cost for sex to an extent that it becomes advantageous over the asexual strategy.

For the case where there is an equal number of sperm and egg-producing organisms, sexual replication will only outcompete asexual replication for large organisms if $ p, \beta < 1/2 $.  This suggests that a sexual replication strategy employing distinct sperm and egg gametes will outcompete an asexual strategy for large organisms if the replication fidelity is sufficiently low, and if the fitness penalty for having a partially defective genome is sufficiently large.  Given that genome length increases with organismal complexity, and given that the genome encodes for an interconnected network of biochemicals, cells, and organs that are crucial for organismal survival, it is likely that the conditions given above hold as $ N \rightarrow \infty $ (though admittedly, we do not have actual data to verify this statement).

\section{Discussion}

\subsection{The evolution of distinct gametes, distinct sexes, and the sex ratio}

The results of this paper suggest that a distinct-gamete sexual replication strategy leads to the selection for sexual replication over asexual replication in larger organisms.  Furthermore, if the replication fidelity is sufficiently low, and if the fitness penalty for having a partially defective genome is sufficiently high, then the fitness benefit for sexual replication can even be shown to overcome the two-fold cost for sex.  

In the context of sexual replication via broadcast fertilization, distinct sperm gametes may have evolved as ``parasitic" gametes defecting from an egg-producing strategy.  The idea is that, in a population consisting entirely of egg-producing organisms, an organism that produces smaller gametes in greater numbers will have a selective advantage, since it can fertilize many eggs for the same cost of producing one egg.  This results in an evolutionary pressure for evolving small gametes that can be produced in great numbers, followed by a co-evolutionary pressure for evolving large gametes that contain the necessary ``yolk" material for allowing the fertilized egg to develop into an immature organism.

Broadcast fertilization is the earliest sexual replication strategy in complex organisms.  This potentially explains why organisms that do not replicate via broadcast fertilization nevertheless employ distinct sperm and egg gametes:  The distinct-gamete strategy proved advantageous for this type of fertilization mechanism, which essentially ``locked" this strategy into subsequent evolutionary lines.  While it is in principle possible for an organismal line to have evolved directly from asexual replicators to sexual replicators via mating, it may be that the body plan and behaviors necessary for mating require an organism more complex than one that sexually replicates via broadcast fertilization.  As a result, it may be far more likely that an organismal line that evolves from asexual replicators to sexual replicators via mating, does so via an intermediate that replicates sexually via broadcast fertilization.

Alternatively, it should be noted that the same evolutionary pressures driving the formation of distinct sperm and egg gametes for broadcast fertilizers also apply to mating organisms.  Therefore, there may be a selective pressure for anisogamy even for a sexual replication strategy that involves mating.  The key difference between the selective pressure for anisogamy in the case of broadcast fertilization, versus the selective pressure for anisogamy in the case of mating, is that in the former case anisogamy is necessary for maintaining a selective advantage for the sexual strategy itself.  In the case of mating, where the time cost associated with haploid fusion is negligible, anisogamy outcompetes other sexual replication strategies, but is not necessary for maintaining a selective advantage over asexual replication.

With distinct sperm and egg gametes, it is possible to have distinct sexes, where one sex, the male, produces only sperm, while another sex, the female, produces only eggs.  The advantage of such a strategy is that since each sex can focus on producing only one type of gamete, the overall efficiency of gamete production is increased.  The disadvantage of such a strategy is that replication is dependent on the mating of a male and a female, and so can have a large fitness penalty if the contact frequency between males and females is too low.  In such cases, it may be advantageous for the organisms to be hermaphrodites, and to be capable of engaging in self-fertilization when necessary.  Indeed, the disadvantage in having distinct sexes has led to many organismal lines preserving a minimal ability for asexual replication, when necessary (it has recently been discovered that female sharks can reproduce asexually when not in contact with any males for a sufficiently long period of time).  In other organismal lines, the organisms can change sexes so as to maintain an appropriate ratio of males to females to make the sexual strategy the preferred one.

It is now established that mammals are the only class of organisms known to be obligately sexual, with distinct male and female sexes \cite{Sharks}.  Presumably, mammals are sufficiently complex that the costs associated with sexual replication are outweighed by its benefits, to an extent that maintaining both sexual and asexual replication pathways, as well as an ability to switch between them, simply incurs a fitness penalty due to the costs involved.

A major issue in evolutionary biology is to understand why the sex ratio in many populations is close to 
$ 1:1 $.  The generally accepted theory holds that the sex ratio is approximately $ 1:1 $, because it is an Evolutionarily Stable Strategy \cite{Charnov:82}.  The basic argument is that, although there are many organisms in which most of the males of the population never mate, it is nevertheless advantageous for half of an organism's offspring to be male, since the few male offspring that do mate will mate with many females and father numerous children.  Furthermore, the production of excess males allows for competition between individual males to claim a group of females.  This competition, if not too costly, provides a natural mechanism to select for beneficial genes that are then passed on to offspring.  

While this paper does not focus on the issue of the sex ratio, we nevertheless wish to bring up two possible alternative explanations for the $ 1:1 $ sex ratio.  In doing so, we should first note that one has to be careful in stating that a $ 1:1 $ sex ratio is an ESS, since in certain kinds of organisms, such as ants and bees, the sex ratio is far from $ 1:1 $ (a colony consists of a single queen, numerous sterile female workers, and few male drones that mate with the queen and fertilize her eggs).

Nevertheless, in certain contexts, a $ 1:1 $ sex ratio may prove advantageous.  For example, in an environment where the population density is low, a $ 1:1 $ sex ratio maximizes the contact rate between males and females, and therefore minimizes the fitness penalty associated with a distinct-sex strategy.  To see this, note that, in a population of $ N $ organisms where the fraction of males is $ x $, the number of males is $ x N $ and the number of females is $ (1 - x) N $, so that the contact rate between males and females is proportional to $ x (1 - x) N^2 $, a quantity that is maximized when $ x = 1/2 $.  Given that larger organisms typically have smaller population sizes, it may be that species that maintain an approximately $ 1:1 $ sex ratio do so because they evolved along organismal lines with relatively small populations.

It is also possible that the $ 1:1 $ sex ratio is a consequence of the fact that certain organisms have specific sex-determining chromosomes, e.g. ``X" and ``Y".  Since each organism is produced by one sperm and one egg, on average one would expect half of all offspring produced to be male, and half female.  

The use of specific sex-determining chromosomes may not necessarily lead to an optimal sex ratio.  However, this method for sex determination may be a relatively simple one in certain environments, so that it might be more robust and less costly than other methods.  As a result, while other sex-determining strategies may lead to more optimized sex ratios, the additional costs they entail exceed the fitness benefit that they provide, so that such strategies are at a net disadvantage.

\subsection{Group selection versus individual selection}

In this paper we have relied on the use of the mean fitness of a population to determine the replication strategy that is advantageous in a given parameter regime.  This approach is known as the {\it group selection} approach, because it assumes that genes that are beneficial for the population as a whole are the ones that will be selected.

Strictly speaking, because an individual organism is the reproducing agent, it is not necessarily true that a strategy that benefits the group is the one that benefits the individual.  Indeed, one of the central points of game theory is that individuals acting in their own self-interest can often engage in behaviors that are detrimental to the group as a whole.  The Prisoner's Dilemma is an excellent example of this.  Biologically relevant examples include the emergence of cancer in multicellular organisms, viral evolution, and possibly even the formation of neural pathways that lead to addictive behaviors \cite{Maynard-Smith:82}.

In general, if the individual organisms in a group are competing for limited resources, so that one individual increases its fitness at the expense of the other organisms, then the group selection approach will not correctly predict what strategies will be selected.  However, if individual organisms can maximize their fitness without adversely affecting the fitness of other organisms, then the group selection approach will correctly predict what strategies will be evolutionarily selected.

In the case of the sexual replication models being considered in this paper, the group selection approach is expected to be valid, because the production of sperm, although it may arise as a form of defection from an investment in egg-production, nevertheless serves to reduce the time cost for sex.  As a result, an organism that produces sperm gametes is not necessarily increasing its own fitness at the expense of other organisms.  On the contrary, a female that invests in the production of haploid eggs that are fertilized by the sperm pays a small time cost for sex.  By contrast, a female that produces haploid eggs that fuse with other haploid eggs pays a large time cost for sex.  As a result, although the former female produces eggs half as quickly as the latter female, the former female's eggs are fertilized much more quickly, and therefore her overall reproduction rate is higher.

Of course, this argument assumes that sperm competition is negligible, or that the fitness benefit of sperm competition outweighs the cost.  Once this no longer holds, then sperm competition will lead to a co-evolutionary dynamics that will result in a reduction of the overall population's mean fitness.  However, as long as this reduction is not so large so as to make asexual replication the preferred replication strategy, the sexual replication strategy will still dominate. 

\section{Conclusions and Future Research}

This paper analyzed the evolutionary dynamics associated with three distinct replication strategies for multicellular, sporulating organisms:  (1)  A purely asexual strategy; (2) A sexual strategy employing identical gametes; (3) A sexual strategy employing distinct sperm and egg gametes.  Under the assumption that the sexual populations replicate via broadcast fertilization, we found that the distinct-gamete strategy is necessary for maintaining a preference for sexual replication as organism size increases.    

As was mentioned earlier in this paper, previous studies exploring the selective advantage for a distinct-gamete strategy have focused on the preference for a distinct-gamete sexual replication strategy over an identical-gamete sexual replication strategy.  This paper, by contrast, argues that a distinct-gamete strategy is necessary for maintaining the selective advantage for the sexual replication strategy itself.
We believe that this is an important result, for a theory explaining the preference for one sexual replication strategy over another does not explain why sexual replication should exist in the first place.

The analysis of this paper relies on the assumption that complex organisms must replicate by producing comparatively large eggs.  As a result, the necessity for producing eggs cannot be dependent on the selective advantage for the distinct-gamete sexual replication strategy itself.  Otherwise, if large eggs were not necessary for organismal viability, then it would make much more sense for organisms to replicate by releasing microscopic single-celled spores.  However, according to the analysis in \cite{Tannenbaum:07}, in this case the asexual pathway would be the advantageous one, and so there would be no selective pressure for evolving any kind of sexual replication strategy.

The models considered in this paper are highly simplified, and so future research should involve developing more realistic replication models.  Some of the features that should be considered are:  (1)  More realistic fitness landscapes, derived from organismal genomes consisting of multiple genes and more than two chromosomes.  Furthermore, the assumption that a single mutation renders a single gene or genome region defective needs to be more closely examined.  (2)  Gamete-release cycles, whereby the organisms do not release their gametes continuously, but rather store up their gametes for a certain period and then release them rapidly and in large quantities.  Presumably, such a strategy can lead to high gamete densities, resulting in high gamete contact rates and rapid fertilization.  This in turn could lead to a more rapid decrease in the cost for sex with increasing organism size than our model predicts.  (3)  Sexual replication models involving explicit mating between organisms, either via external fertilization or internal fertilization.  (4)  Mobile sperm gametes, and eggs that produce pheromones to attract sperm.  (5)  Mortality.

Finally, part of the difficulty in understanding the evolution and maintenance of sex is that there is a wide variety of sexual and mixed sexual-asexual strategies employed by different species, making it difficult to develop a single model for analyzing the selective advantage for sex that is applicable to all organisms.  By studying the time and energy costs, as well as the fitness benefits, of the replication strategies employed by different organisms, it may be possible to obtain a better understanding both of the conditions under which sexual replication is an advantageous strategy, as well as the reasons why organisms adopt the replication strategies that they do.  

\begin{acknowledgments}

This research was supported by a Start-Up Grant from the U.S.-Israel Binational Science Foundation, and by an Alon Fellowship from the Israel Science Foundation.

\end{acknowledgments}

\begin{appendix}

\section{The Dependence of the Mean Fitness $ \bar{\kappa}(t = \infty) $ on $ p $ in the Asexual Replication Model}

We wish to prove that there exists a $ p_{crit} \in [\alpha \beta, \beta] $ such that $ \kappa_{a, 1} > \kappa_{a, 2} $ for $ p \in (p_{crit}, 1] $ and $ \kappa_{a, 1} < \kappa_{a, 2} $ for $ p \in (0, p_{crit}) $.  
We will do this in three steps:

\begin{enumerate}
\item We will prove that $ \kappa_{a, 1} > \kappa_{a, 2} $ for 
$ p \in (\beta, 1] $.
\item We will prove that $ \kappa_{a, 1} < \kappa_{a, 2} $ for $ p \in (0, \alpha \beta) $.
\item We will prove that there exists a unique $ p_{crit} \in [\alpha \beta, \beta] $ such that
$ \kappa_{a, 1} > \kappa_{a, 2} $ for $ p \in (p_{crit}, \beta] $ and $ \kappa_{a, 1} < \kappa_{a, 2} $ for $ p \in [\alpha \beta, p_{crit}) $.
\end{enumerate}

Note that all three statements together complete the proof.

To prove the first step, consider the function $ f(x, y) \equiv -x + \sqrt{x^2 + xy} $.  Note that $ f $ is an increasing function of $ y $ for $ x, y > 0 $.  Note also that,
\begin{equation}
\frac{\partial f}{\partial x} = \frac{x + \frac{y}{2} - \sqrt{x^2 + xy}}{\sqrt{x^2 + xy}}
\end{equation}

Now, for $ x, y > 0 $, $ (x + y/2)^2 = x^2 + xy + y^2/4 > x^2 + xy \Rightarrow x + y/2 > \sqrt{x^2 + xy} $,
so that $ \partial f/\partial x > 0 $ for $ x, y > 0 $, and hence $ f $ is also an increasing function of $ x $ for $ x, y > 0 $.

Therefore, given $ x_1, x_2 $ such that $ 0 < x_1 < x_2 $ and $ y_1, y_2 $ such that $ 0 < y_1 < y_2 $, we have $ f(x_1, y_1) < f(x_2, y_2) $.  

If we set $ x_1 = \alpha $, $ x_2 = 1 $, $ y_1 = 4 (\omega_{vv}/\kappa_{vv}) \beta p $, $ y_2 = 4 (\omega_{vv}/\kappa_{vv}) p^2 $, then for $ p > \beta $ we have $ 0 < x_1 < x_2 $ and $ 0 < y_1 < y_2 $, and so,
\begin{eqnarray}
&  &
\frac{\kappa_{a, 1}}{\kappa_{vv}} = \frac{1}{2}(-1 + \sqrt{1 + 4 \frac{\omega_{vv}}{\kappa_{vv}} p^2}) 
\nonumber \\
&  &
> 
\frac{1}{2}(-\alpha + \sqrt{\alpha^2 + 4 \frac{\omega_{vv}}{\kappa_{vv}} \alpha \beta p}) = 
\frac{\kappa_{a, 2}}{\kappa_{vv}}
\nonumber \\
&  &
\Rightarrow
\kappa_{a, 1} > \kappa_{a, 2}
\end{eqnarray}
thereby proving the first step.

To prove the second step, we will prove the equivalent statement that $ \kappa_{a, 1}/\kappa_{vv} - \kappa_{a, 2}/\kappa_{vv} < 0 $ for $ p \in (0, \alpha \beta) $.

To this end, define, for given $ p $, $ \alpha $, $ \beta $, the function $ g(\lambda) $, via,
\begin{equation}
g(\lambda) = -(1 - \alpha) + \sqrt{1 + \lambda p^2} - \sqrt{\alpha^2 + \lambda \alpha \beta p}
\end{equation}
and note that $ \kappa_{a, 1}/\kappa_{vv} - \kappa_{a, 2}/\kappa_{vv} = (1/2) g(\lambda = 4 (\omega_{vv}/\kappa_{vv})) $.  Therefore, if we can show that $ g(\lambda) < 0 $ for $ p \in (0, \alpha \beta) $, then this will establish that $ \kappa_{a, 1}/\kappa_{vv} - \kappa_{a, 2}/\kappa_{vv} < 0 $ for $ p \in (0, \alpha \beta) $. 

Now, note that $ g(0) = 0 $ and that,
\begin{equation}
\frac{d g}{d \lambda} = \frac{1}{2} p (\frac{p \sqrt{\alpha^2 + \lambda \alpha \beta p} - \alpha \beta \sqrt{1 + \lambda p^2}}{\sqrt{(1 + \lambda p^2)(\alpha^2 + \lambda \alpha \beta p)}})
\end{equation}

We claim that $ d g/d \lambda < 0 $ for $ p \in (0, \alpha \beta) $.  To see this, note that,
\begin{eqnarray}
&  &
\frac{d g}{d \lambda} < 0
\nonumber \\
&  &
\Leftrightarrow
p \sqrt{\alpha^2 + \lambda \alpha \beta p} < \alpha \beta \sqrt{1 + \lambda p^2}
\nonumber \\
&  &
\Leftrightarrow
p^2 \alpha^2 + \lambda \alpha \beta p^3 < \alpha^2 \beta^2 + \lambda \alpha^2 \beta^2 p^2
\nonumber \\
&  &
\Leftrightarrow
\alpha^2 (\beta^2 - p^2) + \lambda \alpha \beta p^2 (\alpha \beta - p) > 0
\end{eqnarray}
This last statement holds for $ p \in (0, \alpha \beta) $, since this implies that $ 0 < p < \alpha \beta \leq \beta $.

Therefore, the identity $ g(0) = 0 $ and the inequality $ g'(\lambda) < 0 $ for $ p \in (0, \alpha \beta) $ together imply that $ g(\lambda) < 0 $ for $ p \in (0, \alpha \beta) $ and $ \lambda > 0 $, and so the second step has been proved.

By continuity, the first and second statements imply that $ \kappa_{a, 1} \geq \kappa_{a, 2} $ for $ p = \beta $, and $ \kappa_{a, 1} \leq \kappa_{a, 2} $ for $ p = \alpha \beta $.

To prove the third and final step, we define $ \lambda = 4 (\omega_{vv}/\kappa_{vv}) $, so that,
\begin{eqnarray}
&  &
\kappa_{a, 1} = \kappa_{a, 2}
\nonumber \\
&  &
\Leftrightarrow
-1 + \sqrt{1 + \lambda p^2} = -\alpha + \sqrt{\alpha^2 + \lambda \alpha \beta p}
\nonumber \\
&  &
\Leftrightarrow
\sqrt{1 + \lambda p^2} - \sqrt{\alpha^2 + \lambda \alpha \beta p} = 1 - \alpha
\nonumber \\
&  &
\Leftrightarrow
\lambda p^2 + \lambda \alpha \beta p + 2 \alpha = 
2 \sqrt{(1 + \lambda p^2) (\alpha^2 + \lambda \alpha \beta p)} 
\nonumber \\
&  &
\Leftrightarrow
\lambda p (p - \alpha \beta)^2 - 4 \alpha (1 - \alpha) (\beta - p) = 0
\end{eqnarray}

Now, when $ p = \alpha \beta $, then $ \lambda p (p - \alpha \beta)^2 - 4 \alpha (1 - \alpha) (\beta - p) =
-4 \alpha \beta (1 - \alpha)^2 \leq 0 $, while when $ p = \beta $, then $ \lambda p (p - \alpha \beta)^2 - 4 \alpha (1 - \alpha) (\beta - p) = \lambda \beta^3 (1 - \alpha)^2 \geq 0 $.  

Therefore, by the Intermediate Value Theorem, there exists a $ p_{crit} \in [\alpha \beta, \beta] $ such that $ \lambda p (p - \alpha \beta)^2 - 4 \alpha (1 - \alpha) (\beta - p) = 0 \Rightarrow \kappa_{a, 1} = \kappa_{a, 2} $.  

To show that $ p_{crit} $ is unique, we differentiate $ \lambda p (p - \alpha \beta)^2 - 4 \alpha (1 - \alpha) (\beta - p) $ to obtain $ \lambda (p - \alpha \beta)^2 + 2 \lambda p (p - \alpha \beta) + 4 \alpha (1 - \alpha) > 0 $ for $ p > \alpha \beta $.  

Therefore, $ \lambda p (p - \alpha \beta)^2 - 4 \alpha (1 - \alpha) (\beta - p) $ is an increasing function for $ p > \alpha \beta $, and hence $ p_{crit} $ must be unique.

Now, if $ \kappa_{a, 1} \leq \kappa_{a, 2} $ for some $ p \in (p_{crit}, \beta] $, then since $ \kappa_{a, 1} \geq \kappa_{a, 2} $ for $ p = \beta $, it follows that there exists some $ p \in (p_{crit}, \beta] $ such that
$ \kappa_{a, 1} = \kappa_{a, 2} $, contradicting the uniqueness of $ p_{crit} $.  Therefore, $ \kappa_{a, 1} > \kappa_{a, 2} $ for $ p \in (p_{crit}, \beta] $.

Similarly, we can show that $ \kappa_{a, 1} < \kappa_{a, 2} $ for $ p \in [\alpha \beta, p_{crit}) $, completing the proof of the third statement.

With all three statements proved, our claim is established.

\section{The effect on the cost for sex if the volume of eggs, sperm, and immature organisms is considered}

If the volume of eggs, sperm, and immature organisms is taken into account, then the system volume will increase at a greater rate than predicted by the models considered in this paper, which assumed that only the adult organisms dictated the total volume of the system.  This will lead to an increased cost for sex than predicted in this paper.  As a result, the asexual strategy will still outcompete the identical-gamete strategy as organism size increases.

However, because the distinct-gamete strategy outcompetes the asexual strategy as organism size increases, it is possible that the increased cost for sex associated with egg, sperm, and immature organism volume will change this result.  

If each cell has a volume $ \nu $, then the total volume occupied by the sexually replicating population is,
\begin{equation}
V = (n_{am, vv} N + n_{ai, vv} f_i N + n_{e, v} f_e N + n_{s, v}) \nu
\end{equation}
so that,
\begin{equation}
\tilde{\rho} \equiv \frac{n_{am, vv} + f_i n_{ai, vv} + f_e n_{e, v} + \frac{n_{s, v}}{N}}{V} = \frac{1}{N \nu}
\end{equation}
where $ f_i $ denotes the average ratio of the number of cells in the immature organisms to the number of cells in the adult, and $ f_e $ denotes the ratio of the number of cells that the eggs contain enough material to produce to the number of cells in the adult.

It is readily shown that,
\begin{eqnarray}
\tilde{\rho} 
& = & 
\frac{n_{am, vv}}{V}(1 + f_i x_{ai, vv} + f_e x_{e, v} + \frac{1}{N} x_{s, v}) 
\nonumber \\
& = &
\rho (1 + f_i x_{ai, vv} + f_e x_{e, v} + \frac{1}{N} x_{s, v}) 
\end{eqnarray}
so that,
\begin{equation}
\rho = \frac{\tilde{\rho}}{1 + f_i x_{ai, vv} + f_e x_{e, v} + \frac{x_{s, v}}{N}}
\end{equation}

As a result, if we re-work the distinct-gamete equations, we obtain an identical set of steady-state equations, except that the cost for sex becomes $ (\kappa_{vv}/(\gamma \tilde{\rho})) (1 + f_i x_{ai, vv} + f_e x_{e, v} + (x_{s, v}/N)) $.  This is still $ \kappa_{vv}/(\gamma \rho) $, except that now $ \rho $ is dependent on the steady-state population ratios.

If the steady-state solution is given by the solution when there is no cost for sex, then the following equalities and inequalities hold:
\begin{eqnarray}
&  &
x_{ai, vv} = \frac{\bar{\kappa}(t = \infty)}{\kappa_{vv}} \leq 2 (1 - \lambda) \frac{\omega_{e, vv}}{\kappa_{vv}} p
\nonumber \\
&  &
x_{e, v} = 0
\nonumber \\
&  &
x_{s, v} \leq \frac{2 \lambda \omega_{s, vv} p}{\bar{\kappa}(t = \infty) + \kappa_{d, s}} \leq
\frac{2 \lambda \omega_{s, vv} p}{\bar{\kappa}(t = \infty)} 
\nonumber \\
&  &
=
\frac{4 \lambda \omega_{s, vv} p}{\kappa_{vv} (-1 + \sqrt{1 + 4 \frac{\omega_{e, vv}}{\kappa_{vv}} p})}
\end{eqnarray}

As $ N \rightarrow \infty $, $ \omega_{e, vv} $ and $ \kappa_{vv} $ both scale as $ N^{-\alpha} $, so $ x_{ai, vv} $ scales less rapidly than $ N^{0} $, and $ x_{s, v} $ scales less rapidly than
$ N^{\beta + \alpha} $, so that $ x_{s, v}/N $ scales less rapidly than $ N^{\beta + \alpha - 1} $.  If $ \beta = 1 - \alpha $, then $ x_{s, v}/N $ scales less rapidly than $ N^{0} $.

Therefore, as $ N \rightarrow \infty $, the steady-state solution when there is no cost for sex is such that the factor $ 1 + f_i x_{ai, vv} + f_e x_{e, v} + x_{s, v}/N $ does not scale more rapidly than $ N^{0} $, so that $ \kappa_{vv}/(\gamma \rho) $ scales no more rapidly with $ N $ as it does in the model considered in this paper.  Since this scaling is such that the steady-state solution as $ N \rightarrow \infty $ is the one when there is no cost for sex, we obtain that the steady-state solution when there is no cost for sex is the self-consistent one as $ N \rightarrow \infty $.  As a result, even if we consider egg, sperm, and immature organism volume, the steady-state solution in our distinct-gamete model is unchanged, and so the conclusions we have drawn from the models considered in this paper are unchanged as well.

\section{A simple mathematical model to justify the necessity for eggs}

In this section, we develop a simple mathematical model that illustrates the necessity for egg production in complex organisms.  We assume that, until a multicellular organism has reached a minimal size consisting of $ f N $ cells, where $ N $ is the total number of cells in the adult and $ f \in (0, 1) $ is a size fraction, the organism has a death rate characterized by a first-order constant $ \kappa_D $.  While this assumption is overly simplistic, it reflects the fact that an organism employing a differentiated, multicellular survival strategy must reach a minimal level of development in order to properly function.

If the adult organism releases spores with enough material to form an organism with $ f' N $ cells, where $ 0 < f' \leq f $, then, following the arguments developed earlier in this paper, we have that the time it takes to produce a single spore is given by,
\begin{equation}
\tau_{spore} = \frac{f'}{\eta} N^{\alpha}
\end{equation}
so that, if $ \omega_{vv}(f') $ denotes the spore production rate when the spore contains enough material to produce $ f' N $ cells, then $ \omega_{vv}(f') = (\eta/f') N^{-\alpha} $.

Now, the initial spore grows from an organism with $ f' N $ cells to an organism with $ f N $ cells.  To determine the characteristic growth time, denoted $ \tau_{grow} $, we note that if $ n $ denotes the number of cells in the organism, then we have,
\begin{equation}
\frac{d n}{d t} = \eta n^{1 - \alpha}
\Rightarrow
\tau_{grow} = \frac{1}{\eta} \frac{f^{\alpha} - f'^{\alpha}}{\alpha} N^{\alpha}
\end{equation}

Since the organisms decay with a first-order rate constant $ \kappa_D $ until they reach a size of $ f N $ cells, and since it takes the newly released spores a time $ \tau_{grow} $ to reach the size of $ f N $ cells, the fraction of newly released spores that reach a size of $ f N $ cells is given by $ e^{-\kappa_D \tau_{grow}} $, so that the effective production rate of immature organisms is given by,
\begin{equation}
\omega_{vv, eff}(f', f) = \omega_{vv}(f') e^{-\kappa_D \tau_{grow}} =
\frac{\eta}{f'} N^{-\alpha} e^{-\frac{\kappa_D}{\eta} \frac{f^{\alpha} - f'^{\alpha}}{\alpha} N^{\alpha}}
\end{equation}

Note that $ \omega_{vv, eff}(f', f)/\omega_{vv}(f) = (f/f') \exp[-(\kappa_D/\eta) ((f^{\alpha} - f'^{\alpha})/\alpha) N^{\alpha}] $, a quantity that goes to $ 0 $ as $ N \rightarrow \infty $.

Therefore, this model suggests that as organism size increases, it makes sense to produce relatively large eggs to maximize fitness.  Of course, the arguments presented above could be used to justify producing eggs that contain enough material to produce the full-sized organism.  What this model neglects is the cost to the fitness of the parent by investing heavily in a given offspring.  Taking such considerations into account would lead to an optimal egg size that is at some intermediate size. 

Despite the simplicity of our model, we nevertheless believe that it illustrates the basic reason for a replication strategy based on the production of eggs in complex organisms (and for parental care in the most complex organisms).

\end{appendix}

\end{document}